\documentclass[english,aps,prl,letter,floatfix,superscriptaddress,twocolumn]{revtex4}
\usepackage[T1]{fontenc}
\usepackage[latin1]{inputenc}
\usepackage{amsmath}
\usepackage{graphicx}
\usepackage{amssymb}

\usepackage{ae,aecompl}
\usepackage{subfigure}

\usepackage{babel}

\begin{document}

\title{Do Binary Hard Disks Exhibit an Ideal Glass Transition?}

\author{Aleksandar Donev}

\affiliation{\emph{Program in Applied and Computational Mathematics}, \emph{Princeton
University}, Princeton NJ 08544}

\affiliation{\emph{PRISM}, \emph{Princeton University}, Princeton NJ 08544}

\author{Frank H. Stillinger}

\affiliation{\emph{Department of Chemistry}, \emph{Princeton University}, Princeton
NJ 08544}

\author{Salvatore Torquato}

\email{torquato@electron.princeton.edu}

\affiliation{\emph{Program in Applied and Computational Mathematics}, \emph{Princeton
University}, Princeton NJ 08544}

\affiliation{\emph{PRISM}, \emph{Princeton University}, Princeton NJ 08544}

\affiliation{\emph{Department of Chemistry}, \emph{Princeton University}, Princeton
NJ 08544}

\affiliation{\emph{Princeton Center for Theoretical Physics, Princeton University},
Princeton NJ 08544}

\begin{abstract}
We demonstrate that there is no ideal glass transition in a binary
hard-disk mixture by explicitly constructing an exponential number
of jammed packings with densities spanning the spectrum from the accepted
{}``amorphous'' glassy state to the phase-separated crystal. Thus
the configurational entropy cannot be zero for an ideal amorphous
glass, presumed distinct from the crystal in numerous theoretical
and numerical estimates in the literature. This objection parallels
our previous critique of the idea that there is a most-dense random
(close) packing for hard spheres {[}\emph{Torquato et al, Phys. Rev.
Lett., 84, 2064 (2000)}{]}.
\end{abstract}
\maketitle

\newcommand{\Cross}[1]{\left|\mathbf{#1}\right|_{\times}}

\newcommand{\CrossL}[1]{\left|\mathbf{#1}\right|_{\times}^{L}}

\newcommand{\CrossR}[1]{\left|\mathbf{#1}\right|_{\times}^{R}}

\newcommand{\CrossS}[1]{\left|\mathbf{#1}\right|_{\boxtimes}}

\newcommand{\V}[1]{\mathbf{#1}}

\newcommand{\M}[1]{\mathbf{#1}}

\newcommand{\D}[1]{\Delta#1}

\newcommand{\sV}[1]{\boldsymbol{#1}}

\newcommand{\sM}[1]{\boldsymbol{#1}}

\newcommand{\grad}{\boldsymbol{\nabla}}

Understanding the glass transition in dense or supercooled liquids
remains one of the challenges of condensed matter physics. In particular,
considerable effort has been directed at identifying the cause of
the dramatic slowdown of the dynamics in the vicinity of the kinetic
glass transition, as evidenced in a decrease of the diffusion coefficient
and an increase in relaxation times. One possibility is that a thermodynamic
transition different from the usual liquid-solid transition underlies
the kinetic one. One scenario originally suggested by Adam and Gibbs
\cite{IdealGlass_AdamGibbs} relates the slow diffusion to a vanishing
of the number of alternative configurations available to the liquid,
leading to an \emph{ideal thermodynamic glass transition} when the
liquid has no choice but to remain trapped in one of few glassy configurations.
An important basic assumption in these considerations is that crystalline
configurations, which are thermodynamically favored, are kinetically
inacessible and therefore the liquid is restricted to exploring {}``amorphous''
configurations. In particular, the term amorphous has become implicitly
attached to the term glass, and crystalline configurations have been
assumed to be qualitatively different from glassy ones. In this Letter,
we study a specific model glass former, namely, a binary hard disk
mixture, and show that, for this model, the presumed {}``ideal glass''
is in fact a phase-separated crystal, and that that there is no special
amorphous (random) state, but rather a continium of states from the
most disordered one to the most ordered one \cite{Torquato_MRJ}.

An inherent-structure formalism was proposed by Stillinger and Weber
and has since been used extensively in the analysis of the thermodynamics
of supercooled liquids \cite{Inherent_Structures}. The {}``inherent-structures''
of hard-particle systems are in fact (collectively) \emph{jammed packings}
\cite{Jamming_g2}, which are mechanically stable packings where the
particles are trapped in a static configuration despite thermal or
external agitation. For soft-particle systems, an essential quantity
in this thermodynamic analysis is the number of distinct energy minima
(basins) with a given energy per particle. For hard-particle systems
this becomes the number of distinct jammed packings $N_{g}(\phi_{J})=\exp\left[Ns_{c}(\phi_{J})\right]$
with jamming packing fraction (density) $\phi_{J}$, where $s_{c}(\phi_{J})$
is the \emph{configurational entropy} (degeneracy) per particle. It
is assumed that the liquid remains in the vicinity of these basins
for long periods of time, jumping from one basin to another as it
explores the available configuration space. Denser packings are favored
in terms of their free-volume; the most favored one being the crystal
of density $\phi_{\textrm{max}}$. However, it is reasonable to assume
that the degeneracy $s_{c}(\phi_{J})$ decreases with increasing $\phi_{J}$.
The liquid achieves minimum free energy by trading off degeneracy
for free volume, so that at a given density $\phi$ it predominantly
samples glasses with jamming density $\hat{\phi}_{J}(\phi)$. The
conjectured ideal glass state corresponds to the point where the number
of available basins becomes subexponential, that is, $s_{c}(\phi_{J}^{\textrm{IG}})=0$.
At densities above an ideal glass transition density $\phi_{\textrm{IG}}$,
defined via $\phi_{J}(\phi_{\textrm{IG}})=\phi_{J}^{\textrm{IG}}$,
the liquid becomes permanently trapped in the ideal glass state. A
crucial unquestioned assumption has been that $\phi_{J}^{\textrm{IG}}<\phi_{\textrm{max}}$,
i.e., that there is a gap in the density of jammed states between
the amorphous and crystal ones. We will explicitly show that this
assumption is flawed for the binary hard-disk mixture we study, and
suggest that this is the case in other similar models, contrary to
numerous estimates for $\phi_{J}^{\textrm{IG}}$ in the literature
\cite{IdealGlass_HS2D_Speedy,ConfigurationalEntropy_Glass_B,ConfigurationalEntropy_Glass_C}.

Previous simulations have cast doubt on the existence of ideal glass
transitions in hard-particle systems \cite{GlassTransition_Crystallization,GlassTransition_Binary}.
In particular, it has already been suggested that the slope of $s_{c}(\phi_{J})$
at $\phi_{J}^{\textrm{IG}}$ dramatically affects the location of
the presumed transition; an infinite slope shifts the transition to
zero temperature \cite{Model_Landscapes}. Questions have also been
raised about the validity of extrapolations into temperature/density
regions that are inacessible to accurate computer simulations \cite{BinaryGlass_DOS_MC},
as well as the impact of finite-size effects \cite{IdealGlass_Equilibration_Critique}.
In this Letter, we present clear evidence that the concept of an ideal
glass transition is flawed for distinctly different reasons. Specifically,
for our model, we \emph{explicitly} construct an exponential number
of jammed packings with jamming densities $\phi_{J}$ that span from
the accepted {}``amorphous'' state with $\phi_{J}^{g}\approx0.84$
to that of the crystal, with $\phi_{\textrm{max}}=\pi/\sqrt{12}\approx0.91$,
thus clearly showing that the configurational entropy cannot be zero
for the hypothetical most-dense amorphous (ideal) glass distinct from
the crystal. This objection is in the same spirit as the critique
of the concept of random close packing (RCP) raised by one of us in
Ref. \cite{Torquato_MRJ}, namely, that there is a continuous tradeoff
between disorder (closely linked to degeneracy) and density, making
the definition of a most-dense random packing ill-defined. Instead,
Ref. \cite{Torquato_MRJ} replaces RCP with the maximally random jammed
(MRJ) state, i.e., the most disordered of all jammed states.

Here we study a binary mixture of disks with a third (composition
$x_{B}=1/3$) of the disks having a diameter $\kappa=1.4$ times larger
than the remaining two thirds ($x_{A}=2/3$). Bidisperse disk packings
with this aspect ratio and $x_{A}=x_{B}=1/2$ have been studied %
\footnote{Our choice of composition is closer to the estimated eutectic point
for disk mixtures with $\kappa=1.4$ than the commonly-used $x_{A}=x_{B}=1/2$,
and also leads to equal area fractions of the large and small disks.%
} as model glass formers \cite{BinaryGlass_Disks}. For this $\kappa$,
it is believed that the high-density phase is a phase-separated crystal.
Our free-energy calculations predict that at the freezing density
$\phi_{F}\approx0.775$, a crystal of predominantly large particles
should start precipitating from the liquid mixture, i.e., systems
in true thermodynamic equilibrium should crystallize just like the
equivalent monodisperse system. Nucleation is however kinetically
strongly suppressed due to the need for large-scale diffusion of large
disks toward the nucleus \cite{GlassTransition_Binary}, and in fact,
we have not observed crystallization even in simulations lasting tens
of millions of collisions per particle well above the estimated freezing
density.

The quantity $N_{g}(\phi_{J})$ has recently been estimated via explicit
enumeration for binary mixtures of relatively small numbers of hard
disks \cite{Disk_RCP_OHern}. These studies have observed an approximately
Gaussian $N_{g}(\phi_{J})$ that is peaked at a density $\phi_{\textrm{MRJ}}\approx0.842$,
interpreted to correspond to the MRJ state for this system. Such a
Gaussian $N_{g}(\phi_{J})$ corresponds to an inverted parabola for
$s_{c}(\phi_{J})$, as has been assumed in previous studies of the
thermodynamics of binary disk glasses \cite{IdealGlass_HS2D_Speedy}.
For large systems, such enumeration is not yet possible and thermodynamics
has been used to obtain estimates of $s_{c}(\phi_{J})$, namely, it
is estimated as the difference between the entropy (per particle)
of the liquid $s_{L}(\phi)$ and the entropy of the {}``glass''
$s_{g}(\phi)$, $s_{c}\left[\hat{\phi}_{J}(\phi)\right]=s_{L}(\phi)-s_{g}(\phi)$.
Here $s_{L}$ is obtained via thermodynamic integration of the \emph{equilibrium}
liquid equation of state (EOS) from the ideal gas limit, while $s_{g}$
is defined as the entropy of the system constrained to vibrate around
a single basin with jamming density $\phi_{J}$, without the possibility
of particle rearrangements. There is significant ambiguity in definining
these constraints; however, at least in the truly glassy region, the
system is typically spontaneously constrained (jammed) by virtue of
a very slow rearrangement dynamics, so that $s_{g}$ can be defined
reasonably precisely.

We have calculated $s_{g}(\phi)$ using a collision-driven molecular
dynamics (MD) algorithm that will be described in detail in a future
publication \cite{Event_Driven_HE}. For hard disks, it is very similar
to the tether method of Speedy \cite{IdealGlass_HS2D_Speedy}. The
particles are restricted to remain inside hard-wall cells in the vicinity
of their initial configuration. The cells are initially about twice
as large as a particle (in linear dimension), which allows particle
vibrations but restricts particle rearrangements, and has been assumed
to define $s_{g}$. The cells then shrink in size slowly during the
course of the MD run, and at the end they become disjoint, leading
to a non-interacting system of particles with analytically-known free
energy. The work done to shrink the cells then gives the initial entropy
of the glass. Closely related methods have previously been used to
calculate configurational entropy \cite{IdealGlass_HS2D_Speedy,ConfigurationalEntropy_Glass_B},
with similar, though less accurate results. For soft-particle glasses,
an alternative method is to use the harmonic approximation to the
vibrational entropy at an energy minimum as an estimate of $s_{g}$
\cite{IdealGlass_HS2D_Speedy,ConfigurationalEntropy_Glass_B}. Both
methods are rigorous only in the high-density (jamming) or $T\rightarrow0$
limit, and are approximate for liquids, so the quantitative results
at low $\phi$ should be interpreted with caution.

\begin{figure}
\begin{center}\includegraphics[%
  width=0.95\columnwidth,
  keepaspectratio]{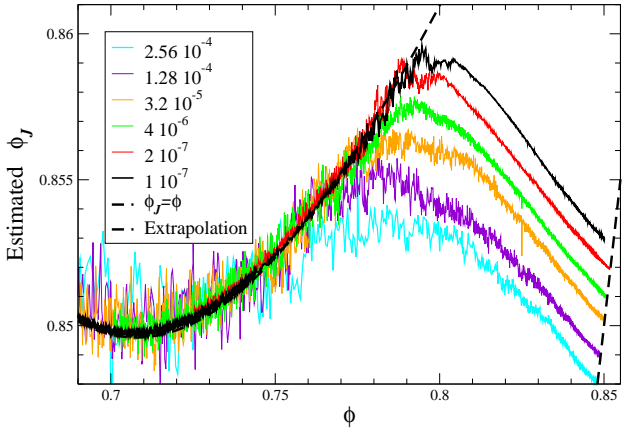}\end{center}

\caption{\label{phi_c.HS.2D.bi.PRL}The equation of state $\tilde{\phi}_{J}(\phi)$
for $N=4096$ disks as observed by compressing a liquid with different
expansion rates $\gamma$ (see legend).}
\end{figure}

The calculation of the true equilibrium liquid equation of state (EOS)
is not possible inside the glassy region with conventional simulation
methods, especially for large system sizes \cite{BinaryGlass_DOS_MC,IdealGlass_Equilibration_Critique,GlassEquilibration_Difficulty}.
In our MD algorithm, we produce glasses by starting with a low-density
liquid and growing the particle diameters at a growth rate $\gamma=dD/dt\ll1$
\cite{Jamming_g2}, for a very wide range of compression rates $\gamma$.
Instead of looking directly at the reduced pressure $p=PV/NkT$, we
assume that a free-volume EOS \cite{Jamming_g2} holds approximately
and estimate the jamming density $\hat{\phi}_{J}(\phi)$ from the
instantaneous pressure\[
\tilde{\phi}_{J}=\frac{\phi}{1-2/p(\phi)}.\]
In the jamming limit ($p\rightarrow\infty$), we get that $\tilde{\phi}_{J}=\phi=\phi_{J}$,
and close to jamming $\tilde{\phi}_{J}\approx\phi_{J}$, and thus
it is much more useful to plot $\tilde{\phi}_{J}(\phi)$ instead of
$p(\phi)$. This is evident in Fig. \ref{phi_c.HS.2D.bi.PRL}, where
we show $\tilde{\phi}_{J}(\phi)$ for a range of $\gamma$'s. We see
that fast compressions fall out of equilibrium at lower kinetic glass-transition
densities $\phi_{g}$, and that the nonequilibrium glassy EOS is very
well described by an empirical $\tilde{\phi}_{J}=(1+\alpha)\phi_{J}-\alpha\phi$,
where $\alpha\approx0.133$, over a wide range of $\phi>\phi_{g}$.
It is also clear that even for the slowest compressions $\phi_{g}\approx0.8$,
so that equilibrating the liquid in reasonable time is not possible
beyond this {}``kinetic glass-transition'' density. Very long MD
runs, with as many as $50$ million collisions per particle, have
failed to equilibriate our samples at a fixed $\phi=0.8$, and in
fact very different microstructures all remained stable for very long
periods of time.

\begin{figure}
\begin{center}\includegraphics[%
  width=0.95\columnwidth,
  keepaspectratio]{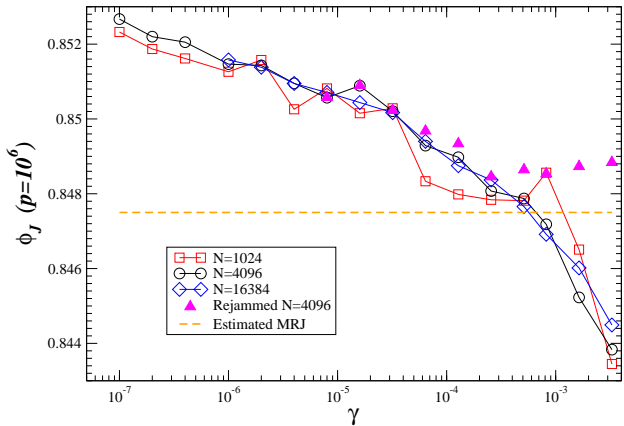}\end{center}

\caption{\label{phi_J.HS.2D.bi}Final jamming density$\phi_{J}$ for different
compression rates $\gamma$, with and without additional relaxation
to ensure a truly jammed packing has been reached.}
\end{figure}

The final jamming densities of the glasses compressed at different
rates are shown in Fig. \ref{phi_J.HS.2D.bi}. Note that slower compressions
consistently yield denser packings with no hints of the existence
of a \emph{densest} glass. Fast compressions produce packings that
are not truly jammed \cite{Jamming_g2} and subsequent relaxation
of these systems increases the density to around $\phi_{J}\approx0.847$.
This behavior of our hard-disk systems is closely related to the observation
that supercooled liquids sample saddle points with the saddle index
diminishing only below the temperature where even the slowest cooling
schedules fall out of equilibrium \cite{GlassTransition_Saddles},
i.e., the kinetic glass transition temperature. Observations similar
to those in in Fig. \ref{phi_J.HS.2D.bi} have already been made for
systems of soft particles, e.g., the energy of the lowest inherent-structure
sampled has been shown to continuously decrease for slower cooling
\cite{Inherent_Structures}.

\begin{figure}
\begin{center}\includegraphics[%
  width=0.95\columnwidth,
  keepaspectratio]{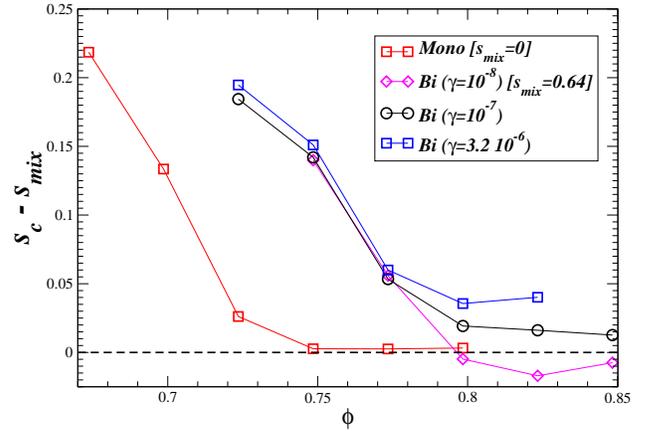}\end{center}

\caption{\label{dA_SCO.HS.2D}Estimated $s_{c}(\phi)-s_{\textrm{mix}}$ for
monodisperse and bidisperse systems of $N=4096$ disks, as obtained
from (sufficiently slow) compressions with a range of $\gamma$'s.}
\end{figure}

The measured $s_{c}(\phi)$ for the different glass compressions are
shown in Fig. \ref{dA_SCO.HS.2D}. For comparison, the results for
a slow compression of a monodisperse system are also shown and the
entropy of mixing $s_{\textrm{mix}}=x_{A}\ln x_{A}+x_{B}\ln x_{B}$
has been subtracted from $s_{c}$. It is seen that for the monodisperse
case $s_{c}-s_{\textrm{mix}}$ ($s_{\textrm{mix}}=0$ in this case)
becomes very nearly zero after the liquid freezes (around $\phi\approx0.7$),
indicating a continuous or a very mildly discontinuous liquid-solid
phase transition. More interesting is the fact that $s_{c}-s_{\textrm{mix}}$
also becomes nearly zero for the binary glasses around the kinetic
phase transition (around $\phi\approx0.8$). This important observation
has not been made before. It means that the estimated number of packings
that the liquid samples near the glass transition is very close to
$s_{\textrm{mix}}$, which is also the entropy of the uncorrelated
ensemble of discrete states in which a fraction $x_{A}$ of the particles
is chosen to be large and the remaining particles are chosen to be
small. It is interesting to observe that the parabolic fit to $s_{c}(\phi_{J})$
from the work in Ref. \cite{Disk_RCP_OHern}, if constrained to equal
the mixing entropy at the maximum, passes through zero at $\phi\approx0.9$,
much higher than the extrapolation in \cite{IdealGlass_HS2D_Speedy}
and close to the crystal jamming density. We note that all measurements
of $s_{c}$ in the literature that we are aware of are above or close
to $s_{c}$ near the kinetic glass transition, and all estimates of
the zero crossing of $s_{c}$ are based on extrapolations beyond this
point without numerical support \cite{IdealGlass_HS2D_Speedy,ConfigurationalEntropy_Glass_B,ConfigurationalEntropy_Glass_C}.

Such extrapolations are flawed and, in fact, an exponential number
of amorphous jammed packings exist over the whole density range from
that accepted as the MRJ density $\phi_{\textrm{MRJ}}\approx0.84$
to that of the phase-separated crystal $\phi_{\textrm{max}}\approx0.91$.
Lower-density jammed packings also exist \cite{Disk_RCP_OHern}; however,
they do not have thermodynamic significance and thus our simulations
do not generate them. In our simulations we observe that higher $\phi_{J}$
implies microsegregation in the form of increased clustering of the
large particles. This has been most vividly demonstrated in Ref. \cite{BinaryGlass_Disks}.
This observation suggests that one can generate denser packings by
artificially encouraging clustering. 

\begin{figure}
\begin{center}\includegraphics[%
  width=0.45\columnwidth,
  keepaspectratio]{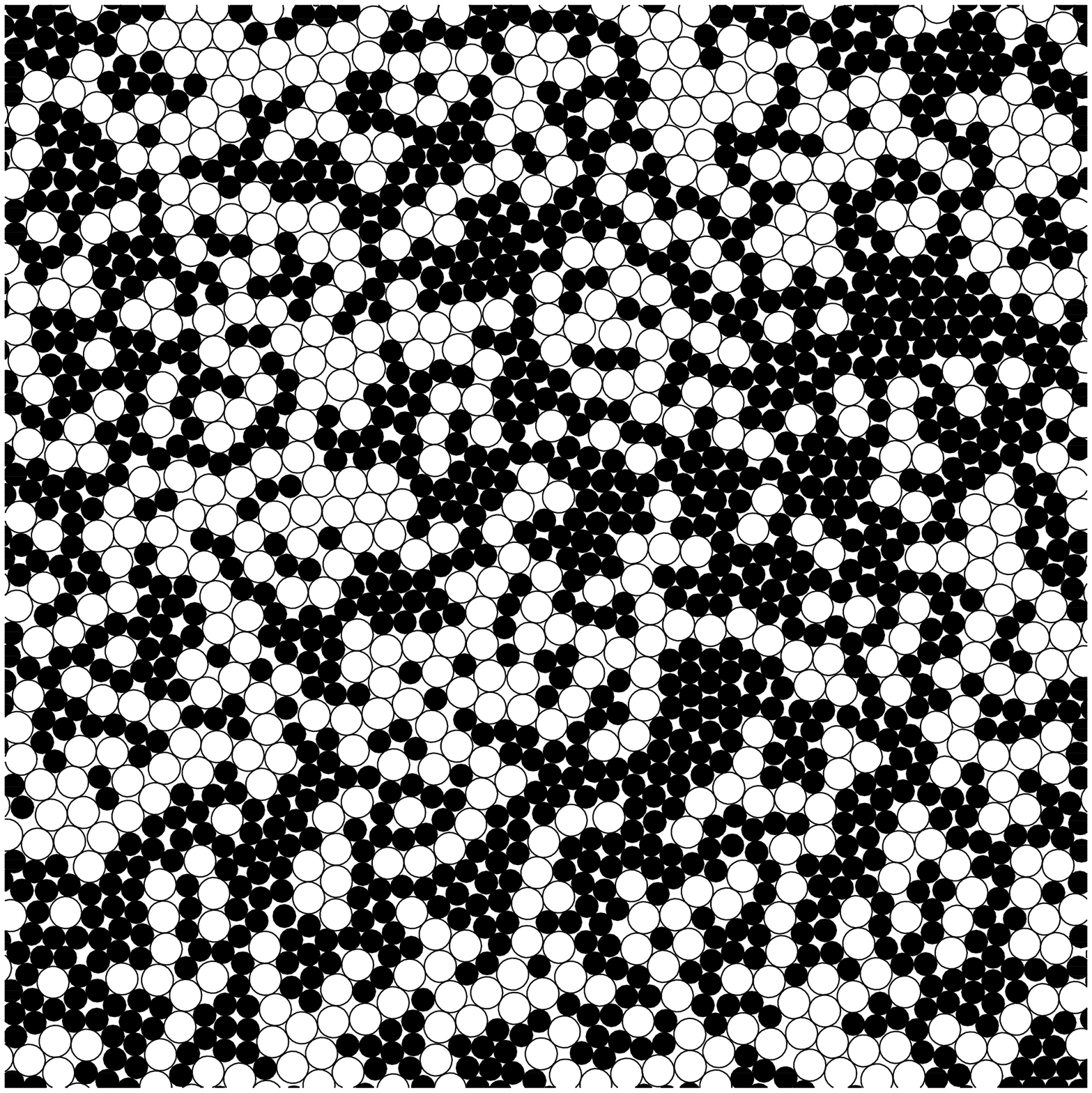}\hspace{0.25cm}\includegraphics[%
  width=0.45\columnwidth,
  keepaspectratio]{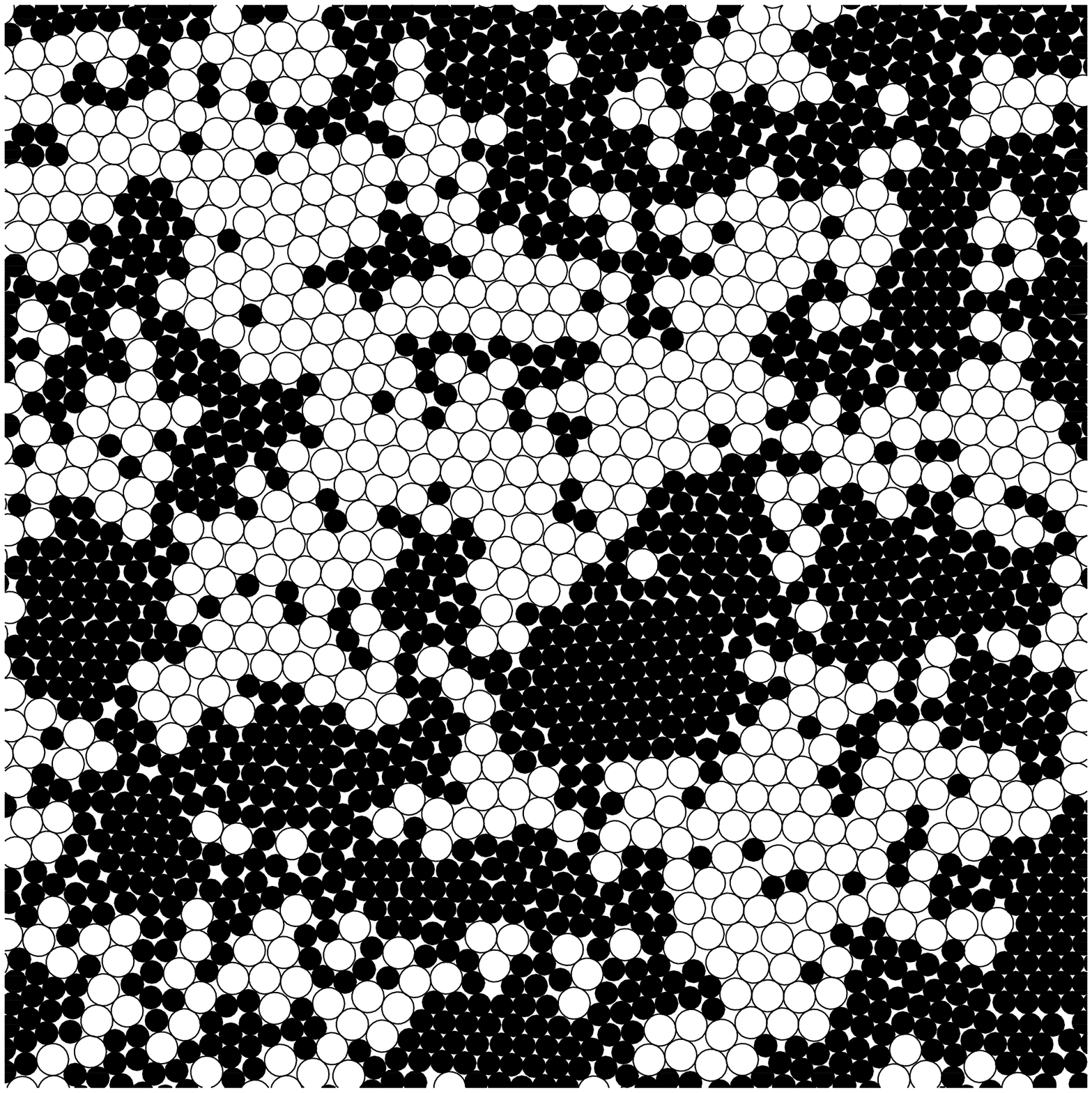}\end{center}

\caption{\label{LSD.HS.bi.2D.Matern}The microstructure of a packing without
(left, $\phi_{J}\approx0.846$) and with moderate clustering (right,$\phi_{J}\approx0.850$).}
\end{figure}

To achieve clustering, we start from a monodisperse ($\kappa=1$)
triangular crystal at pressure $p=100$ in which a third of the particles
has been selected as being {}``large''. The large particles then
slowly grow in diameter while the system is kept in (quasi)equilibrium
at a constant (isotropic) pressure using a Parinello-Rahman-like variation
of the MD algorithm \cite{Event_Driven_HE}. When $\kappa=1.4$ we
stop the process and then slowly compress the system to jamming. By
spatially biasing the initial partitioning into large and small disks,
we can achieve a desired level of clustering and higher jamming densities.
For this purpose we use a level-cut of a Gaussian random field (GRF)
with suitably chosen parameters for a flexible family of pair correlation
functions \cite{PhaseSeparated_Entropy}. Figure \ref{LSD.HS.bi.2D.Matern}
illustrates two different jammed packings, one with an uncorrelated
random choice of large disks, and another with correlations encouraging
microsegregation.

\begin{figure}
\begin{center}\includegraphics[%
  width=0.95\columnwidth,
  keepaspectratio]{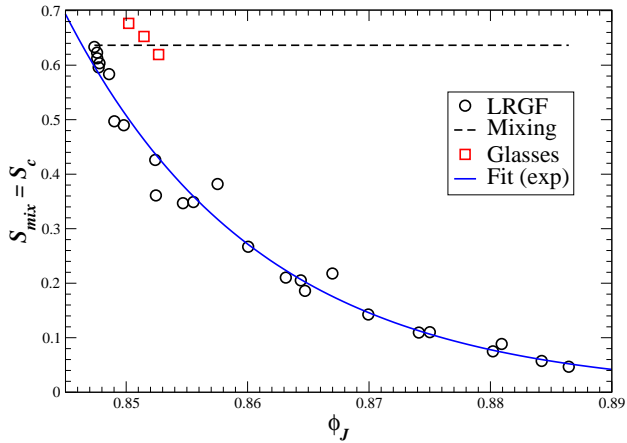}\end{center}

\caption{\label{S_c.phi.matern}The measured degeneracy of packings of $N=4096$
disks obtained by using different parameters of a random Gaussian
field with Matern correlations \cite{PhaseSeparated_Entropy}, as
a function of the jamming density. For comparison, we have shown $s_{c}(\phi=0.825$)
for the three compressions shown in Fig. \ref{dA_SCO.HS.2D}.}
\end{figure}

To determine the configurational entropy (degeneracy) for a given
choice of the GRF parameters, we use a recently-developed algorithm
for obtaining numerical approximations of the entropy (per site) of
lattice systems \cite{Entropy_MarkovExpansion}. The algorithm numerically
measures the probabilities of observing a given configuration, which
in our case is the partitioning of the triangular lattice into large
and small disks, for small windows (we have used $4\times4$ windows).
This can be done by generating sufficiently many GRF realizations
and counting the number of times a given configuration occurs. A Markov
expansion is then used to approximate the entropy per site $s_{c}$.
Figure \ref{S_c.phi.matern} shows our results for $s_{c}$ versus
the jamming density $\phi_{J}$, for a wide choice of the GRF parameters.
The results clearly show that in order to increase $\phi_{J}$ one
must sacrifice degeneracy ($s_{c}$). The figure also shows the first
\emph{measured}, rather than extrapolated, estimate of $s_{c}(\phi_{J})$.
This observed $s_{c}(\phi_{J})$ only goes to zero for the phase-separated
crystal state, rather than the hypothetical amorphous ideal glass
state postulated by extrapolations.

Continuing on work in Ref. \cite{Torquato_MRJ}, we explicitly demonstrated
that the concept of random close packing as the most-dense jammed
amorphous packing is flawed. By trading off degeneracy for density
in a continuous manner, we constructed an exponential number of amorphous
jammed packings with densities spanning the range from the most disordered
to most ordered jammed states. We explicitly calculated, as opposed
to extrapolated, the degeneracy entropy for densities well above that
of the postulated ideal glass transition, and found that the degeneracy
is positive for all {}``amorphous'' states and vanishes only for
the phase-separated crystal. For the maximally random jammed state,
the found a degeneracy entropy very close to the mixing entropy. Furthermore,
our free-energy calculations predict a thermodynamic crystallization
well-below the kinetic glass transition, casting additional doubt
on the search for a thermodynamic origin of the glass transition.
Although the present study focused on the hard-disk binary mixture,
the fundamental principles are general enough to be applicable to
a host of related systems, notably, both mono- and bi-disperse with
hard-core and soft interactions.

\begin{acknowledgments}
This work was supported in part by the National Science Foundation
under Grant No. DMS-0312067.
\end{acknowledgments}

\end{document}